\documentstyle[preprint,prl,aps]{revtex}
\input psfig
\begin{document}
\draft

\preprint{}
\title{Thomas-Fermi Theory of Fullerenes}

\author{Dennis P. Clougherty\cite{byline} and Xiang Zhu}

\address{Department of Physics, University of Vermont, Burlington, VT 
05405}

\date{August 1995}

\maketitle

\begin{abstract}
We study {{\rm C}$_{60}$} with the use of Thomas-Fermi theory.
A spherical shell model is invoked to treat the nuclear potential,
where the nuclear and core charges are smeared out into a shell of 
constant surface charge density. The valence electron distribution and
the electrostatic potential are efficiently computed by integration of
the Thomas-Fermi equation, subject to the shell boundary conditions.
Total energy is numerically calculated over a range of shell radii,
and the mechanical stability of the model is explored, 
with attention to the constraints of Teller's theorem.
The calculated equilibrium radius of the shell is in 
good agreement with experiment.
\end{abstract}

\pacs{PACS Numbers: 31.15.Bs, 36.40.Qv, 31.10.+z}

The highly symmetrical structure of {{\rm C}$_{60}$}
 has motivated geometrical
approximations which have previously been invoked to study electronic 
and optical properties of the molecule. While consideration of the
icosahedral structure of the molecule is necessary for detailed
comparison 
with experiment, previous studies have had success in describing 
some of the properties of {{\rm C}$_{60}$} 
within the continuum approximation, where a system of free electrons 
are constrained to moving on the surface of a 
sphere\onlinecite{ozaki,murthy,gonzales,savina}.

While the peak electron density should be found on the shell, 
electrostatic consideration of the mechanical
stability of the entire system 
requires that a sizeable fraction of the total number of valence 
electrons be {\em inside} the shell. Motivated by this observation, we
study  
here a generalization of the previously considered continuum model: we
allow  
the valence electrons to move in three dimensions in the external
potential 
generated by a spherical shell of constant surface charge density.
We call such an artificial molecule whose nuclear potential has
spherical shell 
symmetry ``spherene.'' 

We treat spherene by the Thomas-Fermi (TF) method. While TF results
are typically rather rough, it is often used to efficiently generate
starting potentials for more exact self-consistent field methods. 
We have had success\cite{zhu}
 using the resulting TF potential in this fashion. Of course, TF
theory has historically had value in its own right.
In this Letter, we 
use the TF results to discuss the stability of 
{{\rm C}$_{60}$}. This is rather
subtle business, as it is well-known that local density methods such
as TF are cases for Teller's theorem\cite{teller} which states that 
molecules in TF theory will unbind. We prove Teller's theorem in the
context of a spherical shell, and we circumvent it by considering the
true point charge distribution in the calculation of the ``nuclear''
energy. 

The approach used here was previously employed by N.~March\cite{march}
to investigate molecules with the form of a central atom with
tetrahedrally 
or octahedrally coordinated ligands, such as in the case of  ${\rm
CH_4}$ and ${\rm SF_6}$. We closely follow March's work, applying it
to the case of {{\rm C}$_{60}$}, where there is no central atom and an
icosahedral arrangement of ``ligands.'' The absence of a central atom
changes the boundary condition of March at the origin. The 
case of {{\rm C}$_{60}$} would seem to 
be  ideally suited to this approach, given its high ``coordination''
number; and the case of an endohedrally-doped fullerene, such as {\rm
La@{{\rm C}$_{60}$}}, is of the exact form considered by March.  

We start with a positively charged spherical shell of radius $R$ and
charge $Ze$. The shell charge arises from the sum of the 
positive nuclear charges
with the core electrons of the constituent atoms. We take for the
case of {{\rm C}$_{60}$}, 
$Z=60$; thus, the valence electrons are the remaining 60
$\pi$-electrons. 
These valence electrons interact with the shell via
a spherically symmetric cut-off Coulomb
potential, given by
\begin{equation}
V_n(\vec r)=\left\{
\begin{array}{ll}
                          {Ze/R}&\ \mbox{$0<r<R$}\\
& \\
                        {Ze/r}&\ \mbox{$r\ge R$}
\end{array}
\right.
\label{cut}
\end{equation}

We can view this model as arising from an approximation of the true
nuclear 
potential where all but the monopole term is neglected. 
The validity of the spherical shell model
can be
examined by expanding the nuclear potential in multipole moments. 
We denote the
location of the i$^{th}$ atom by a radius $R$ and a set of spherical
angles 
$\Omega_i=(\theta_i, \phi_i)$. We center our coordinate system on the
geometric center, and we align our axes with the five-fold and
two-fold axes of the molecule. Thus, for the region external to the
cage,

\begin{eqnarray}
V_n(\vec{r})
&=& \sum_{\ell,m}{1\over {r^{\ell+1}}} \sqrt{4\pi\over2\ell+1}Q_{\ell m}Y^*_{\ell m}(\Omega)
\label{pole}
\end{eqnarray}
$Q_{\ell m}$ is 
the $2^\ell$-pole moment, given by

\begin{equation}
Q_{\ell m}= e {R^\ell} \sqrt{4\pi\over2\ell+1} \sum_i Y_{\ell m}(\Omega_i)
\end{equation}

The summation in Eq.~\ref{pole} 
is only over even $\ell$ as a result of the inversion
symmetry of {{\rm C}$_{60}$}. Furthermore, in general we note that
$Q_{\ell m}$
is non-vanishing only if the spherical irreducible representation,
denoted by $\ell$, when decomposed in terms of the irreducible
representations of $I_h$ contains the trivial (a$_{1g}$) 
representation. 

Thus, after the non-vanishing
monopole moment, the next non-vanishing elements are in $\ell=6$,
followed by $\ell=10$. For {{\rm C}$_{60}$}, 
we need not consider $\ell > 10$, as
the highest lying electron orbital is derived from an $\ell=5$
manifold. $Q_{\ell m}$ is also only non-vanishing for $m=0$ and $\pm 5$.

We estimate the error of neglecting $\ell\ne 0$ terms by evaluating 
 the relevant dimensionless parameters
\begin{equation}
\alpha_{\ell m}=\bigg|\sqrt{4\pi\over 2\ell+1}
{Q_{\ell m}\over R^{\ell} Q_{0 0}}\bigg| 
\end{equation}
We find that 
$\alpha_{6, 0}=0.026$, $\alpha_{6, 5}=0.020$, 
$\alpha_{10, 0}=0.021$, and
$\alpha_{10, 5}=0.034$. As $\alpha_{\ell m} \ll 1$ for $\ell\le 10$,
we conclude that the spherical approximation is reasonable for our
purposes.

We consider the dimensionless TF equation 
without exchange effects at
temperature $T=0$,
\begin{equation}
{d^2\phi \over dx^2}={\phi^{3 \over 2}\over x^{1 \over 2}}
\label{tf}
\end{equation}
$x$ is the distance from the center of the shell in units of 
\begin{equation}
b={1\over 4}\left[{9\pi^2 \over 2Z} \right]^{1 \over 3}a_0
\end{equation}
where $a_0$ is the Bohr radius of hydrogen. $\phi$ is related to the
potential in the usual way
\begin{equation}
V(r)={Ze \over r}\phi(x)
\label{pot}
\end{equation}

Without nuclear charge at the origin, the standard atomic boundary
condition at $x=0$ is altered to $\phi(0)=0$, as the potential is now
finite at the origin. The presence of the shell gives rise to a
discontinuity in the derivative of $\phi$ at the shell. Thus,
\begin{equation}
\phi'(X^-)-\phi'(X^+)={1 \over X}
\label{bc}
\end{equation}
where $X$ is the shell radius in dimensionless units and
differentiation is with respect to $x$. Additionally, $\phi$ itself is
continuous over its domain, and $\phi\to 0$ as $x\to\infty$.   

We obtain numerical solutions to Eq. \ref{tf} subject to the above
boundary conditions for different values of $X$. A variation of the
shooting method\cite{recipes} is used where we choose a trial slope for
$\phi$ at the origin, and we  
integrate outward to the asymptotic region ($x \gg X$). The boundary
condition at infinity is replaced by requiring that $\phi$ vanish at
an outer shell of large radius. The slope is subsequently varied in a
systematic 
way until the boundary condition on the outer shell is satisfied.

The following identities are used as a final check of the numerical
procedures: (1)
conservation of 
electron number requires that $\phi$ satisfy  
\begin{equation}
\int_0^{\infty} \phi^{3 \over 2} x^{1 \over 2} dx=1
\end{equation}
and (2) the virial theorem for shell systems requires that
\begin{equation}
{9 \over 35}\int_0^{\infty} \phi^{5 \over 2}x^{-{1 \over
2}}dx+\int_0^{\infty} {\phi^{3 \over 2}x^{1 \over 2} \over x_>} dx=-2
\left[\phi'(X)-{\phi(X) \over X} \right] 
\end{equation}
where $x_>$ is the larger of $x$ and $X$. The resulting solutions
obey the above relations to an accuracy of better than three parts in
$10^4$. 

\begin{figure}
\psfig{file=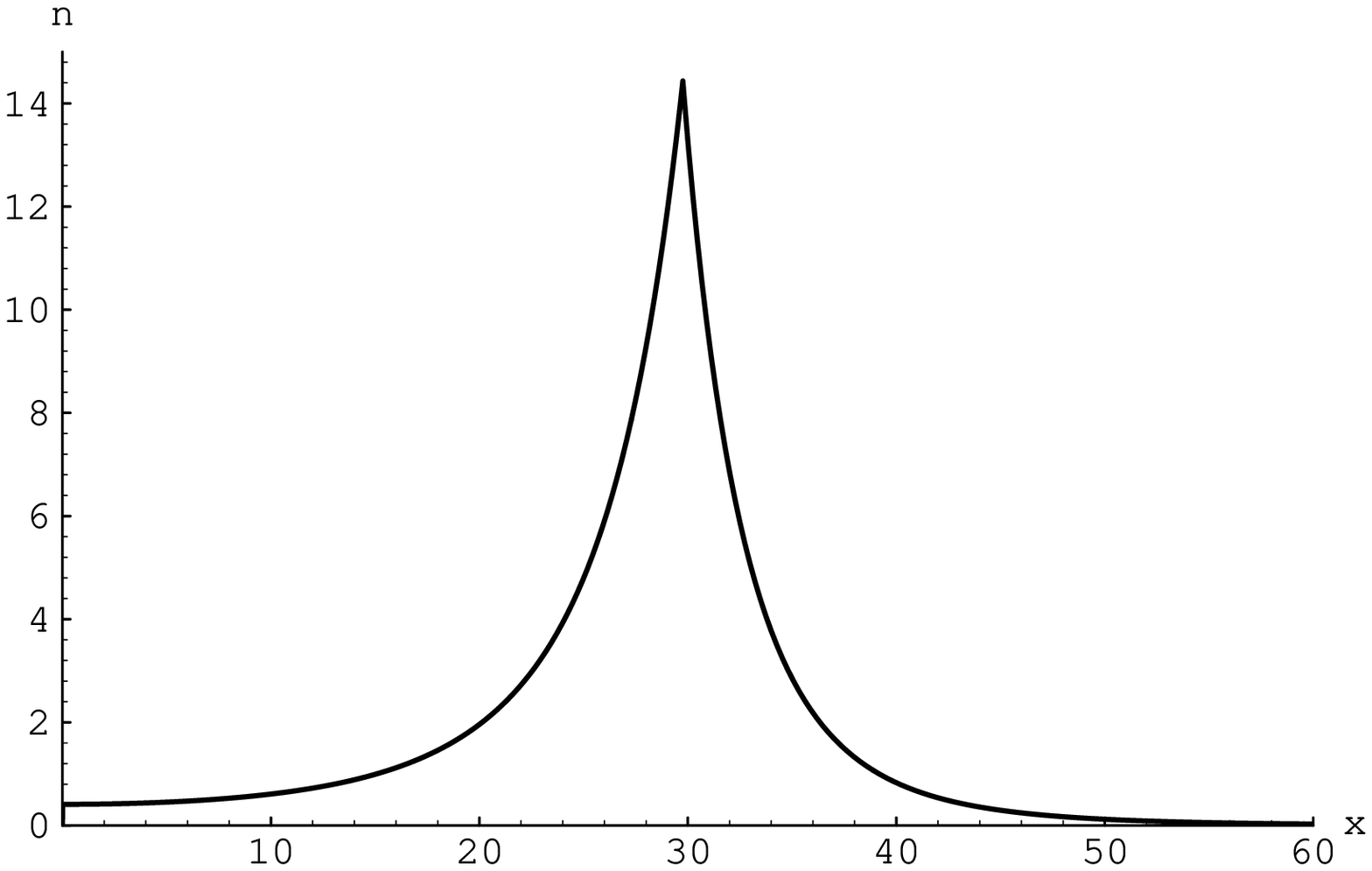,width=5in}
\caption{Electron charge density $n$ (in units of ${{15\over\pi}\times
10^5 b^{-3}}$) vs $x$ for $Z=60$
and $R= 6.73 a_0$.\label{nc60}}
\end{figure}

In Fig.~\ref{nc60}, we show the electron charge density 
$n$ as a function of $x$ obtained from our solution for $\phi$ for
parameters corresponding to those modeling {{\rm C}$_{60}$}, 
where $X=29.7592$
($R=6.73 a_0$) and
$Z=60$. 
While
the potential is found from Eq.~\ref{pot}, $n$ is computed from the
relation
\begin{equation}
n(x)={Z \over 4 \pi b^3} \left[{\phi(x) \over x}\right]^{3 \over 2}
\label{n}
\end{equation}
We note
that $\phi$ (and consequently $n$ and $V$) is strongly peaked at the
shell, in a consistent fashion with the continuum models which
constrain the 
valence electrons to the surface of the shell; however, it is
significant that nearly 43\% of the valence 
electrons are contained inside the shell. We return to this point in
our discussion of stability.

We follow March\cite{march} and conclude that the electronic energy
$E_e$ can be simplified to a form 
requiring only values of $\phi$ and its derivative evaluated just
inside the shell, 
\begin{equation}
E_e={Z^2e^2\over 7bX}\big[4 \phi(X^-)-X \phi'(X^-) -3\big]
\label{e-en}
\end{equation}
For the {{\rm C}$_{60}$} parameters, we find $E_e = -486$ Ry.

Using the Hellman-Feynman theorem, we 
calculate the radial force that the 
electrons exert on the shell, $F_r$: 
\begin{eqnarray}
F_r&=&-{dE_e\over dR}\\
&=&-Ze{dV(R)\over dr}\\
&=&-{Z^2e^2\over b^2X^2}\big[X\phi'(X^-)-\phi(X^-)\big]
\label{force}
\end{eqnarray}

Only electrons in the interior of the shell can exert a force on the
shell, a consequence of Gauss' Law. Hence, the presence of
charge in the interior provides a centripetal, stabilizing force which
opposes the centrifugal self-force of the shell.

From dimensional considerations, we write the self-interaction energy
of the shell as
\begin{equation}
U_n=c{Z^2e^2\over R}
\label{e-shell}
\end{equation}
For the uniform shell, $c={1\over 2}$. 
Within TF theory, Teller's theorem\cite{teller} implies that the
stabilizing force of the 
electrons on the shell is of insufficient magnitude to compensate for
the repulsive self-force of the shell. Thus, there is no finite 
equilibrium radius for the shell for $c={1\over 2}$. We sketch a
proof
specific to the shell.
The proof proceeds by {\em reductio ad absurdum}.

We assume that there exists a finite equilibrium shell radius
$R_0$. Using $c={1\over 2}$, Eqs.~\ref{force} and
Eq.~\ref{e-shell} imply that
\begin{equation}
E(R_0)={3\over 7} ZeV(R_0^-)
\label{eqe}
\end{equation}
Since $V$ is positive for finite $r$, we conclude that $E(R_0)$ is
positive as well. However, the virial theorem states that at
equilibrium
\begin{equation}
E(R_0)= -T
\label{virial}
\end{equation}
where $T$ is the total kinetic energy of the electrons. Since $T$ must
be positive, we conclude from Eq.~\ref{virial} that $E(R_0)$ is
negative. But this is in contradiction to the result of Eq.~\ref{eqe}.
Hence, there is no finite $R_0$.

\begin{figure}
\psfig{file=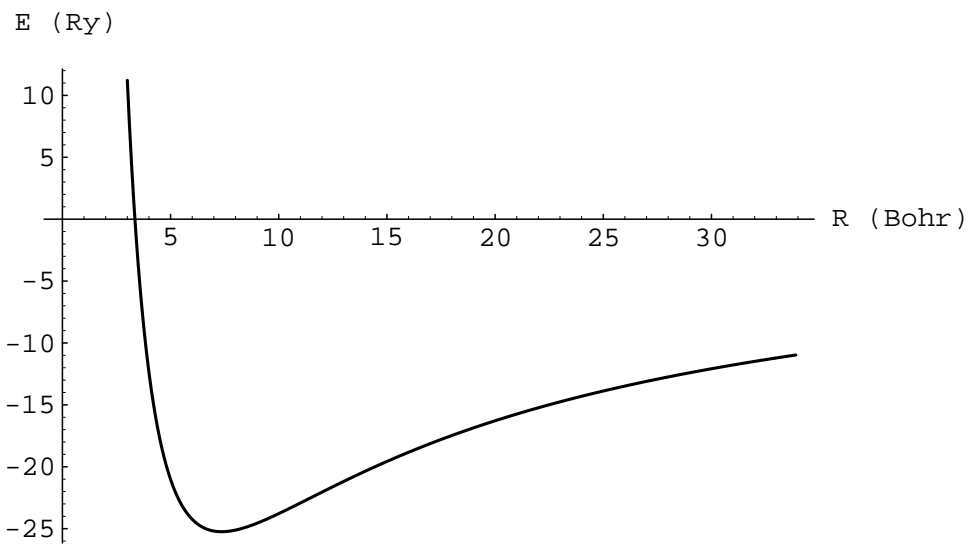,width=5in}
\caption{Total energy $E$ vs shell radius $R$.\label{eqstate}}
\end{figure}

The continuum approximation overestimates the shell self-force. If
one computes the self-force by considering the point structure of the
ions, a stable equilibrium is obtained. For the system of $Z$ ions of 
charge $e$ located on the vertices of a truncated icosahedron of
radius $R$, we find $c\approx 0.4311$. We use this value of $c$ to
compute the total energy of the system, $E=E_e+U_n$, at different
shell radii. The resulting energy curve is plotted in
Fig.~\ref{eqstate}. From the minimum of the curve, we extract an
equilibrium radius, $R_0=7.36 a_0$, which is in good
agreement with the experimental value\cite{pickett} of $6.73 a_0$. 

It is amusing to
observe that at equilibrium, $c$ is precisely the fraction of valence
electrons contained inside the shell. This condition follows from
simple electrostatic considerations.

In addition to giving an equilibrium radius in good agreement with
experiment, 
TF gives a potential which is an
excellent starting potential for more rigorous self-consistent field
techniques. Furthermore, 
there are many enhancements of this method which can be easily
incorporated, such as the inclusion of exchange and correlation and
density gradient corrections to the kinetic energy.

The method can be extended to other fullerene systems of interest. 
It is a simple
matter to treat endohedrally-doped fullerenes
or
positively-charged fullerenes. 
Lastly, by generalizing to finite temperature, equations of state can be
calculated, as was done
previously in the case of atoms\cite{feynman}. 

We thank F. Anderson, F. Ham, and L. Scarfone for valuable
discussions. We thank M. McHenry for providing us with 
I$_h$ group code which we used to generate the {{\rm C}$_{60}$} 
coordinates. Acknowledgment is made to the Donors of The Petroleum
Research Fund, administered by the American Chemical Society, for
support of this research.
This work has also been supported in part by the AFOSR Summer Faculty
Research Program. X.~Zhu also acknowledges support from the UVM
Graduate College.

\vfill\eject

\end{document}